\documentclass[letterpaper]{article} 
\usepackage{aaai25}  
\usepackage{times}  
\usepackage{helvet}  
\usepackage{courier}  
\usepackage[hyphens]{url}  
\usepackage{graphicx} 
\usepackage{natbib}  
\usepackage{caption} 
\usepackage{algorithm}
\usepackage{algorithmic}
\usepackage{booktabs}

\usepackage{newfloat}
\usepackage{listings}
\DeclareCaptionStyle{ruled}{labelfont=normalfont,labelsep=colon,strut=off} 
\floatstyle{ruled}
\newfloat{listing}{tb}{lst}{}
\floatname{listing}{Listing}
\title{Preventing Another Tessa: Modular Safety Middleware For Health-Adjacent AI Assistants}
\author {
    Pavan Reddy\textsuperscript{\rm 1},
    Nithin Reddy\textsuperscript{\rm 2},
}
\affiliations {
    \textsuperscript{\rm 1}The George Washington University\\
    \textsuperscript{\rm 2}Tulane University\\
    pavan.reddy@gwmail.gwu.edu, nreddy1@tulane.edu,
}

\usepackage{bibentry}

\begin{document}

\maketitle

\begin{abstract}
In 2023, the National Eating Disorders Association’s (NEDA) chatbot Tessa was suspended after providing harmful weight-loss advice to vulnerable users—an avoidable failure that underscores the risks of unsafe AI in healthcare contexts. This paper examines Tessa as a case study in absent safety engineering and demonstrates how a lightweight, modular safeguard could have prevented the incident. We propose a hybrid safety middleware that combines deterministic lexical gates with an in-line large language model (LLM) policy filter, enforcing fail-closed verdicts and escalation pathways within a single model call. Using synthetic evaluations, we show that this design achieves perfect interception of unsafe prompts at baseline cost and latency, outperforming traditional multi-stage pipelines. Beyond technical remedies, we map Tessa’s failure patterns to established frameworks (OWASP LLM Top10; NIST SP 800-53), connecting practical safeguards to actionable governance controls. The results highlight that robust, auditable safety in health-adjacent AI does not require heavyweight infrastructure: explicit, testable checks at the last mile are sufficient to prevent “another Tessa,” while governance and escalation ensure sustainability in real-world deployment.
\end{abstract}

\section{Introduction}
AI assistants are increasingly deployed in health and counseling contexts, where even minor errors can result in significant harm. In 2023, the National Eating Disorders Association (NEDA) transitioned its scripted prevention chatbot, “Tessa,” from a supplementary tool to the primary point of contact for individuals seeking help \cite{wired_tessa_suspend_2023}. Although the initial deployment proceeded without issue, public reports later indicated that Tessa was promoting calorie deficits, weigh-ins, and other weight-loss behaviors incompatible with evidence-based eating disorder care. As a result, NEDA suspended the chatbot \cite{guardian_tessa_2023, wired_tessa_suspend_2023}. This incident highlights a key principle: in high-risk domains, content moderation must not be treated as an afterthought, and safety mechanisms should be explicitly engineered as testable, first-class components of the system, rather than relying solely on prompt design.

This paper analyzes the Tessa failure as a concrete case of unsafe assistance in a clinical-adjacent context and proposes a minimal, deployable remedy: a lightweight safety middleware that \emph{ deterministically} enforces health-specific redlines while keeping latency and cost compatible with real-world operations. The design combines (i) a fast lexical gate that blocks obviously risky intents (e.g., calorie targets, weigh-ins, dieting frames) and (ii) an \emph{in-line} large language model (LLM) policy filter that judges and, when allowed, generates responses in a single call. The module emits a strict JSON verdict (\texttt{is\_safe}, violation categories), operates fail-closed, and performs a final numeric/lexical scan of the buffered answer before rendering. Crisis signals trigger escalation and safe-mode responses by construction.

Our contributions are threefold. First, we map the observed breakdowns in Tessa to generalizable safety failures including, but not limited to policy like implementation drift, missing input triage, insecure output handling, etc. and connect them to OWASP LLM Top10 and NIST800-53 controls to make the lessons actionable. Second, we present a modular safety layer that integrates a denylist/regex fast path with an LLM policy filter that produces both an answer (when permissible) and a self-audit verdict, avoiding the latency and cost of two-model pipelines. Third, we prototype this design and demonstrate, on representative interactions, that it intercepts disallowed content while preserving responsiveness, illustrating how modest engineering can lead to safer versions of ``Tessa"

\begin{table*}[t]
\centering
\begin{tabular}{|p{2.5cm}|p{14cm}|}
\hline
\textbf{Date} & \textbf{Event Description} \\
\hline
February 2022 & Tessa debuts on NEDA’s website as a scripted, prevention-focused chatbot operating alongside the human helpline. \cite{Nguyen2023InsiderDisable} \\
March 17, 2023 & Helpline staff win their union election; NLRB certification follows later in the month. \cite{Nguyen2023InsiderNEDA} \\
March 27, 2023 & NLRB certifies the Helpline Associates Union. \cite{guardian_tessa_2023} \\
March 31, 2023 & NEDA tells staff the helpline will be wound down and services will transition to Tessa around June~1. \cite{Wells2023KFFNEDA} \\
May 26, 2023 & Helpline Associates Union publicly warns that replacing trained humans with a chatbot endangers help-seekers. \cite{Nguyen2023InsiderNEDA} \\
May 29, 2023 & Activist Sharon Maxwell posts evidence that Tessa recommends caloric deficits, weekly weigh-ins, and other weight-loss guidance. \cite{Nguyen2023InsiderDisable} \\
May 30, 2023 & NEDA suspends Tessa pending investigation, citing harmful and non-programmed responses. \cite{Wells2023KFFNEDA} \\
June 1, 2023 & National coverage details the helpline’s closure and Tessa’s suspension, with criticism of reliance on a chatbot. \cite{Nguyen2023InsiderNEDA} \\
\hline
\end{tabular}
\caption{Timeline of key public events surrounding NEDA’s ``Tessa'' chatbot.}
\label{tab:timeline}
\end{table*}

\section{Overview and Timeline of Events}

The National Eating Disorders Association (NEDA) introduced the chatbot ``Tessa” as a scripted, rule-based, prevention-oriented resource intended to expand access to support for individuals experiencing body image concerns and eating disorders \cite{wired_tessa_suspend_2023,Wells2023KFFNEDA}. This deployment occurred in parallel with the continued operation of a long-running, human-staffed helpline \cite{Wells2023KFFNEDA,wired_tessa_suspend_2023}. NEDA initially described Tessa as an evidence-based tool that delivered structured content derived from cognitive-behavioral and related therapeutic frameworks \cite{wired_tessa_suspend_2023,Wells2023KFFNEDA}. Early use did not generate public controversy.

In late March 2023, internal labor developments brought the chatbot to broader attention \cite{Nguyen2023InsiderNEDA,Wells2023KFFNEDA}. Following the certification of the helpline staff’s union on March 27, NEDA leadership notified employees that the live helpline would be discontinued, with services transitioning to Tessa \cite{Wells2023KFFNEDA,Nguyen2023InsiderNEDA}. Layoffs were announced at the end of March \cite{Nguyen2023InsiderNEDA,Wells2023KFFNEDA}. Although NEDA denied that the decision was motivated by union activity—citing strategic and operational considerations—the timing prompted criticism \cite{Nguyen2023InsiderNEDA,guardian_tessa_2023}. Observers argued that replacing human-led support with automation in a high-risk mental health setting could compromise the quality of care \cite{Nguyen2023InsiderNEDA,wired_tessa_suspend_2023}.

Public concern intensified in late May 2023, when clinicians and advocacy groups documented instances of Tessa providing responses that contradicted established eating disorder treatment principles \cite{guardian_tessa_2023,wired_tessa_suspend_2023}. Reports indicated that the chatbot offered weight-loss advice and recommended behaviors such as calorie counting and routine weigh-ins—interventions widely considered clinically inappropriate and potentially harmful for this population \cite{guardian_tessa_2023,wired_tessa_suspend_2023}. These findings reinforced earlier ethical concerns raised by former helpline staff and community members, who emphasized that automated systems lack the empathy, contextual awareness, and dynamic risk assessment required to support individuals in psychological distress \cite{Nguyen2023InsiderNEDA,Wells2023KFFNEDA}.

In response to the growing evidence of harm, NEDA suspended Tessa on May 30, 2023, pending further investigation \cite{guardian_tessa_2023,Wells2023KFFNEDA}. The organization acknowledged that the chatbot’s outputs had diverged from the intended, policy-consistent messaging \cite{guardian_tessa_2023}. NEDA stated that Tessa would remain offline until adequate safeguards could be implemented \cite{Wells2023KFFNEDA}. However, the simultaneous closure of the helpline created a service gap, with users redirected to external crisis support resources. This case has since been cited as a cautionary example in the application of AI to mental health care, highlighting the importance of accountability, oversight, and the continued relevance of human judgment in high-risk therapeutic contexts \cite{HarvardChan2023AIHarm}.

\section{Background and Related Work}

Modern AI deployments in healthcare expand access and efficiency but also widen the blast radius of security failures: systems handle protected health information, influence clinical decisions, and expose novel attack surfaces. Core risks of such systems span confidentiality (training-data memorization, model inversion, and membership inference that can leak PHI \cite{carlini2021extracting,fredrikson2015modelinversion,shokri2017membership}), integrity (data poisoning and prompt/tool-injection that redirect model behavior \cite{biggio2012poisoning,greshake2023promptinjection}), and operational safety (insecure output handling, excessive autonomy, brittle configuration) \cite{owasp_llm_top10_proj}. For this readon, the deployment of AI in healthcare must be accompanied by rigorous safety and security methods that systematically mitigate these risks, thereby reducing the likelihood of harm to patients, providers, and institutions.

Research in safety for clinical-adjacent assistants supports layered moderation combining deterministic and model-based checks.

\textbf{\cite{bai2022constitutional}} showed that constitutional prompting can steer models away from unsafe outputs.

\textbf{\cite{ouyang2022training}} showed that instruction tuning with human feedback improves adherence but does not eliminate domain-specific risks.

\textbf{\cite{zheng2023judging}} showed that \emph{LLM-as-a-judge} pipelines can reliably audit and filter model drafts.

\textbf{\cite{madaan2023selfrefine}} showed that self-refinement loops let models critique and revise their own drafts, reducing unsafe generations in single-call workflows. 

On the detection side, \textbf{\cite{hartvigsen2022toxigen}} showed with ToxiGen that toxicity stress tests expose limits of static blocklists and motivate supervised detectors

\textbf{\cite{greshake2023promptinjection}} showed that prompt-injection attacks can bypass naïve policy-as-prompt setups, underscoring the need for deterministic post-generation scanners and mediation boundaries. 

Together, these results justify the hybrid approach used in this paper: a lexical fast path for obvious redlines and an in-line LLM policy filter that emits a strict, fail-closed verdict alongside any response.

Governance frameworks make these technical choices auditable and operational. NIST SP~800-53 (Rev.5) \cite{nist_sp80053r5_2020} provides control families—configuration/change management, incident response, audit and accountability, and supply-chain risk—that map directly onto AI assistant operations (e.g., versioned prompts/policies, rollback plans, crisis handoff, provenance for external models and data) \cite{nist_airmf_2023}. 

The OWASP Top10 \cite{owasp_llm_top10_proj} for LLM Applications identifies common failure modes—prompt injection, insecure output handling, excessive agency, insecure configuration—that align with the Tessa postmortem and motivate concrete mitigations (input mediation, output sanitization, capability scoping, monitoring). For readers new to the area, these frameworks supply the “what to require” (NIST controls) and the “what to test” (OWASP risks) so that health-adjacent assistants implement enforceable safeguards rather than relying on intent alone.

\section{Vulnerabilities and Threat Model}

The failure of NEDA’s “Tessa” providing calorie-restriction and weight-loss guidance to people seeking eating-disorder support  was not merely content-specific—it revealed missing gates, weak change control, and insufficient operational safety. These are the same classes of issues that routinely undermine AI deployments and can be directly aligned to established security taxonomies and controls.

Even though Tessa was not a GenAI chatbot (it was a scripted flow), its security issues map cleanly to GenAI systems: risky inputs passed through ungated paths; outputs were not deterministically filtered against redlines; session memory, escalation, monitoring, and rollback were inadequate. In GenAI, these gaps present as prompt-injection exposure, insecure output handling, excessive agency without human oversight, and brittle supply-chain/change management.

\subsection{Security Issues (Generalized with Tessa Examples and AI Mappings)}
\begin{itemize}
\item \textbf{Policy–implementation drift and weak change control.}
Tessa produced outputs that contradicted its documented redlines, suggesting that content or template changes, as well as vendor updates, were deployed without rigorous review. In generative AI (GenAI) systems, similar drift occurs when prompts, retrieval sources, or tool manifests evolve more rapidly than approval processes and testing, resulting in safety regressions reaching production.

\item \textbf{Missing input triage and risk–intent detection.}
Tessa handled dieting and weight-loss queries within an eating disorder context as standard traffic. In GenAI, the analogous failure is the absence of intent gating and adversarial prompt screening, allowing harmful or manipulative inputs—including paraphrased or injection-based variants—to bypass safeguards.

\item \textbf{Insecure output handling and lack of deterministic redlines.}
Tessa generated calorie targets, weigh-ins, and measurement tracking data without a final-stage blocking mechanism. GenAI systems with unstructured decoding and no post-generation filtering exhibit similar vulnerabilities, leaking restricted content to users or downstream components.

\item \textbf{No escalation pathway or human handoff for crisis signals.}
Signals such as expressions of body dissatisfaction or restrictive dietary goals did not trigger escalation to human experts in Tessa. GenAI systems demonstrate comparable risks when agents operate autonomously on sensitive inputs without deferring to refusal flows or human oversight.

\item \textbf{Stateless moderation and lack of session-level risk accumulation.}
Tessa moderated interactions on a turn-by-turn basis, preventing the accumulation of risk signals across a session. GenAI deployments that disregard prior user flags similarly fail to escalate protective measures, allowing repeated borderline prompts to evade detection.

\item \textbf{Insufficient monitoring, red-teaming, and rollback mechanisms.}
Harmful templates in Tessa were not identified or withdrawn in a timely manner. In GenAI contexts, the absence of shadow evaluations, canary deployments, anomaly detection, and rollback infrastructure enables jailbreaks, retrieval drift, and policy regressions to persist unmitigated.

\item \textbf{Third-party and supply-chain risk without provenance or auditability.}
Vendor-supplied scripts in Tessa operated outside the intended cognitive behavioral therapy (CBT) framework. GenAI architectures that incorporate external models, plugins, datasets, or safety filters require provenance tracking, version control, and audit trails to avoid inheriting opaque and unmanaged risks.

\item \textbf{No automated safe-mode or system-level kill-switch.}
Tessa required manual suspension upon discovery of harm. GenAI systems should instead include automated degradation mechanisms that reduce capabilities or revert to deterministic safe outputs in response to violations or anomalous activity, restoring full function only after remediation.
\end{itemize}

Table~\ref{tab:tessa-llm-mapping} shows a mappings from each Tessa failure pattern to an AI/LLM vulnerability class, with corresponding OWASP Top 10 for LLM applications entries and NIST SP 800-53 control mappings.

\begin{table*}[t]
\centering
\small
\caption{Mapping Tessa security issues to AI/LLM applications, OWASP LLM Top 10 \cite{owasp_llm_top10_proj}, and NIST 800-53 controls\cite{nist_sp80053r5_2020}}
\label{tab:tessa-llm-mapping}
\begin{tabular}{|p{3.5cm}|p{3.5cm}|p{3.5cm}|p{2.5cm}|p{2.5cm}|}
\toprule
\textbf{Security Issue} & \textbf{Tessa Failure Pattern} & \textbf{AI/LLM Vulnerability Class} & \textbf{OWASP LLM Top 10} & \textbf{NIST 800-53 (Rev. 5) Controls } \\
\midrule
Policy–implementation drift \& weak change control & Content/templates diverged from redlines without vetting & Insecure configuration; ungoverned prompt/content drift & LLM10, LLM05 & CM-3, CM-5, SA-10, SA-11, SA-9 \\
Missing input triage and risk-intent detection & Risky dieting intents processed as normal & Absent input validation; prompt/injection exposure & LLM01, LLM10 & SI-10, AC-4, SC-7 \\
Insecure output handling (no deterministic redlines) & Harmful numeric/behavioral guidance delivered & Unvalidated model outputs; unsafe post-processing & LLM02 & AC-4, SI-10 \\
No escalation pathway / human handoff & Crisis signals not routed to humans & Excessive autonomy on sensitive intents & LLM08, LLM09 & IR-4, IR-8, AU-12 \\
Stateless moderation (no session risk memory) & Turn-by-turn checks; no accumulated risk & Context loss across turns; non-escalating safeguards & LLM09, LLM10 & AC-16, AU-12, SI-4 \\
Insufficient monitoring, red-teaming, and rollback & Harm persisted; no canaries/rapid rollback & Lack of continuous evaluation and ops controls & LLM10, LLM05 & CA-7, RA-5, SA-11 \\
Third-party/supply-chain risk without provenance & Vendor changes produced unsafe behavior & Unvetted external components/models/data & LLM05 & SA-9, SA-12, SR-3, SR-5 \\
No automated safe-mode or kill-switch & Manual suspension required post-harm & Missing automated containment and fail-safe & LLM08, LLM10 & IR-4(1), CP-10 \\
\bottomrule
\end{tabular}
\end{table*}

\section{Proposed Safety Module Design}

This safety design is scoped specifically to the failure modes evidenced in the Tessa case. We focus on non-adversarial weaknesses observed there (e.g., policy–implementation drift, missing input/output gates, absent escalation, inadequate monitoring/rollback) and deliberately exclude broader AI security vectors such as prompt injection, data poisoning, model exfiltration, or agent/tool abuse, since Tessa’s failure did not involve a malicious actor. The architecture is modular: the same gating, adjudication, and safe-mode/rollback primitives can be extended to adversarial settings by adding injection-resistant retrieval, content provenance and allowlisting, tool sandboxing, and anomaly/attack detection—without altering the core pipeline.

\subsection{Improved Prompting and Generation Policy}

We enforce strict system-level instructions that explicitly prohibit domain-specific harms, such as weight-loss advice, calorie targets, weigh-ins, BMI coaching, and restrictive-diet framing. Refusal-first behavior is mandated for disallowed intents, with empathetic redirection to appropriate resources. Stylistic constraints include prohibiting numeric calorie or weight guidance and dieting-related terminology. Anti-bypass prompting is integrated to prevent circumvention via user-provided “policies,” role-play scenarios, or embedded instructions. Content embedded in links, quotations, or attachments is explicitly disregarded.

Prompting constitutes only a single layer of control and remains brittle under adversarial conditions, such as obfuscated language (e.g., “c@l0r!es”), indirect requests (e.g., “what helps someone slim down”), long-context drift, and jailbreaking.

\subsection{Input Filtering (Keywords and Classifiers)}  

We filter user queries to block or route high-risk requests before generation. Two complementary strategies are employed:

\begin{itemize}
    \item \textbf{Keyword/Regex Fast Path (Deterministic).}  
    A denylist targets dieting- and eating disorder-related terms as well as numeric patterns. Examples include:
    \begin{verbatim}
(1) \b(lose|cut)\s+weight\b
(2) \b(deficit|restrict|weigh[- ]?
    in|skinfold|BMI)\b
(3) \b\d{2,4}\s*(k?cal|calories)\b
    \end{verbatim}
    Localized variants are maintained, including slang, misspellings, leetspeak, and homoglyph substitutions. This path provides low latency and high explainability. However, it introduces a precision–recall trade-off: overly strict patterns may harm user experience, while lenient ones may miss risky inputs.

    \item \textbf{Classifier Path}  
    A policy classifier processes the full input text. Possible implementations include: (i) using an LLM as a classifier with a few-shot safety rubric; (ii) a lightweight CNN or 1D convolutional model; (iii) a fine-tuned transformer-based model (e.g., RoBERTa or BERT) trained to detect eating-disorder risk and dieting-related intent. Ensemble methods and threshold calibration by locale or domain improve robustness. 
    
    A continuous risk score \(p \in [0,1]\) is defined with thresholds \(t_{\mathrm{lo}} < t_{\mathrm{hi}}\): block the input if \(p \ge t_{\mathrm{hi}}\); allow it if \(p \le t_{\mathrm{lo}}\); and trigger review or clarification if \(t_{\mathrm{lo}} < p < t_{\mathrm{hi}}\).

    For pilot applications, lack of training data constrains the feasibility of (ii) and (iii).
\end{itemize}

\subsection{Output Moderation (Post-Generation Check)}

We enforce final checks on generated text to ensure redlines are upheld at delivery.

\begin{itemize}
    \item \textbf{LLM-based adjudication.} The buffered draft is sent to a judge model prompted with the safety policy. Prefer a different model family or prompting profile than the generator to reduce correlated failure.
    
    \item \textbf{Deterministic scanners.} In parallel, lexical/numeric scans search for calorie counts, deficits, weigh-ins, BMI targets, and restrictive-diet phrases (including covert frames such as “tone,” “shred,” or “cut”).
    
    \item \textbf{Decision.} Content is rendered only if all checks return \texttt{SAFE}. If any return \texttt{UNSAFE} or \texttt{UNCERTAIN}, a standardized refusal is issued or the case is escalated. Adjudications are cached for identical outputs to reduce cost and latency.
    
    \item \textbf{Latency control.} Inference is gated on the first failing signal (parallel judge + scanners) to minimize delay. In streaming responses, generation halts immediately on a failing signal.
\end{itemize}

\begin{figure*}[t]
\centering
\includegraphics[width=\textwidth]{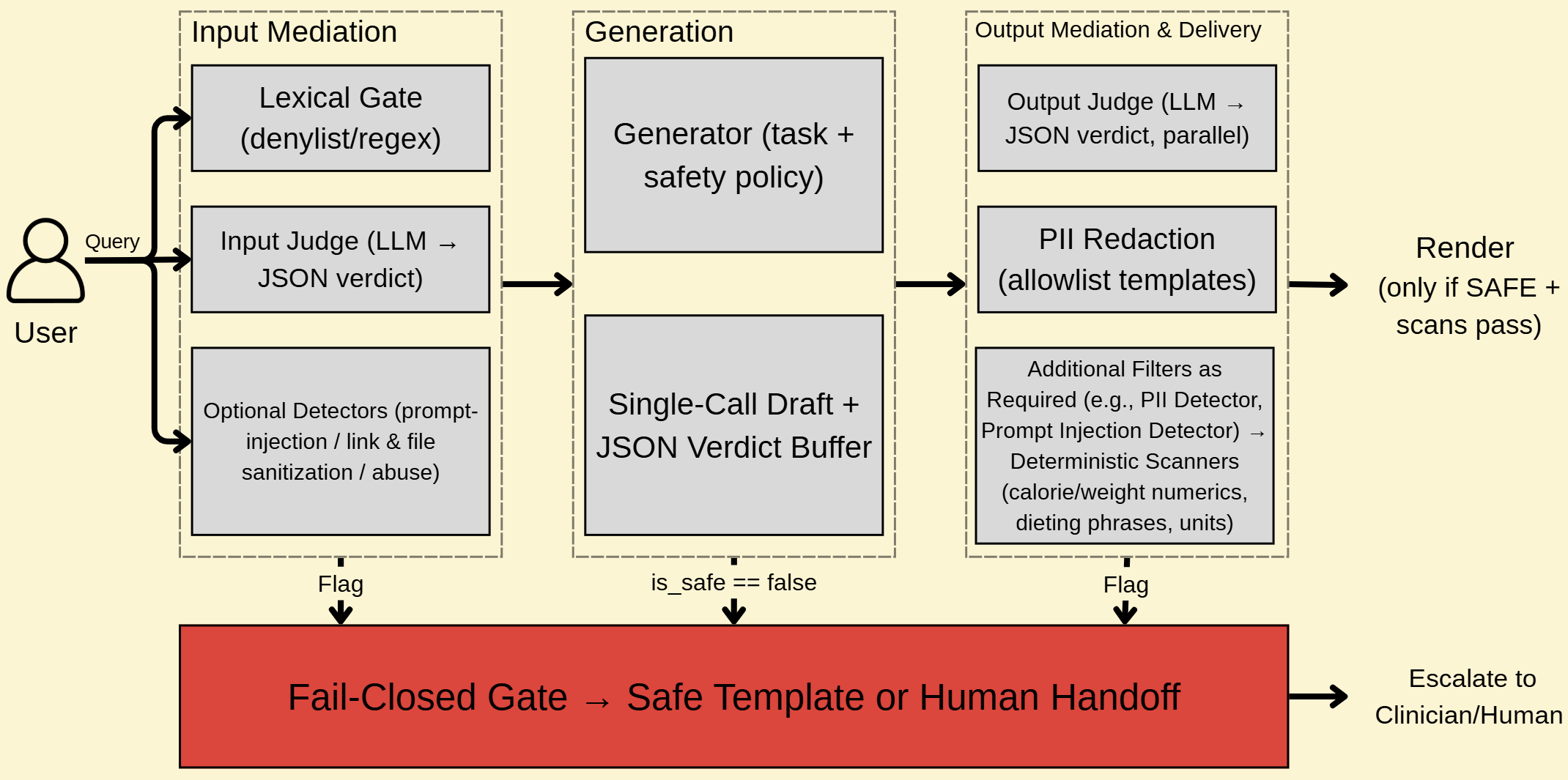}
\caption{Safety middleware pipeline with input mediation, secure generation, output moderation, and a fail-closed gate that escalates unsafe queries to humans.}
\label{fig:safety-pipeline}
\end{figure*}

\subsection{Single-Call JSON Mode (Cost/Latency Mitigation)}

All prior modules introduce additional latency and cost. Our Single-Call JSON Mode mitigates these issues by combining generation and classification into a single call.  

\noindent \textbf{Pattern.} The model is instructed to emit the full answer text \emph{followed by} a strict JSON verdict. Ordering matters: the verdict follows the answer so the classifier has full context. At the API boundary, only the trailing JSON is parsed; the buffered answer is dropped if the verdict is unsafe or unparsable (fail-closed).  

If the LLM identifies a harmful request, it emits no answer (\texttt{null}) and sets \texttt{is\_safe=false}, reducing cost. If unsafe content is generated, the trailing JSON verdict (produced with the benefit of full autoregressive context) enforces safety without requiring a second model call. \\

\noindent\textbf{Illustrative schema for a harmful query (e.g., \texttt{"Give me a 1200 calorie meal plan"}):}
\begin{verbatim}
{"response": null,
 "is_safe": false,
 "violations": [
   "dieting/weight-loss guidance", 
   "calorie targets"
 ]
} 

\end{verbatim}

\noindent\textbf{Illustrative schema for a non-harmful query (e.g., \texttt{"What are some healthy recipes for people with diabetes?"}):}
\begin{verbatim}
{"response": "<assistant_answer_text>",
 "is_safe": true,
 "violations": []
} 

\end{verbatim}

Systems must operate with a fail-closed default, ensuring that no unadjudicated or unparsable content is ever rendered. Determinism at delivery mandates a final numeric or lexical re-scan of the buffered output, even in cases where \texttt{is\_safe=true}. To maintain consistency, any discrepancy between the initial verdict and the deterministic scan should default to refusal or escalation, invoking a safe-mode response. Finally, observability requires comprehensive logging of inputs, decisions, verdicts, and scan features—subject to privacy controls—to enable effective auditing, drift detection, and rapid rollback.

\section{Methodology}
We evaluate the proposed safety modules for an eating disorder (ED) support chatbot under a strict policy that prohibits guidance related to weight loss, dieting, calorie targets or deficits, weigh-ins, measurement logging, BMI coaching, fasting, meal-skipping, or any other content likely to trigger ED-related behaviors. Queries indicating crisis intent, such as self-harm or suicidality, must trigger a refusal response and escalation protocol. The architecture is implemented using SecuRAG \cite{securag}, an open-source security and auditing framework. All the assets and screenshots from SecuRAG, along with audit logs are provided in the github repo \footnote{The prompts, keywords used in the filtering mechanism, system prompts, and the code are provided in \url{https://github.com/pavanreddyml/AAAI-Tessa}. }

\subsection{Environment}
All filters were implemented in a Python~3.11 runtime. For reproducibility, we employed the Google Gemma~2 (2B) model as the generation backbone. The same model, under different prompting configurations, was used for both the Input-Judge and Output-Judge components. While best practice favors a model family distinct from the generator for output adjudication, computational constraints and reproducibility requirements led us to reuse the same model.

\subsection{Dataset}
We constructed a synthetic evaluation set of \textbf{100 prompts}: \textbf{50 malicious} (queries that directly or indirectly seek dieting or ED-unsafe content, including through euphemisms, numeric expressions, obfuscations, and indirect framings) and \textbf{50 safe} (queries focused on body-neutral support, coping strategies, treatment access, and general wellness without weight-loss framing). Labels were known by construction, based on policy criteria. The dataset was generated using GPT-4o-mini via API in JSON mode.  

Prompts were vectorized using the OpenAI \texttt{text-embedding-3-small} model, and near-duplicate entries were removed to maximize coverage of diverse surface forms.

\subsection{Deployment Patterns}
We compare six deployment patterns, holding the assistant’s core tasking constant:
\begin{enumerate}
  \item \textbf{Insecure Zero-Shot Prompting (A)  \label{sec:A}:} A baseline configuration where the model is prompted with only generic tasking instructions and no embedded safety rules, redlines, or gating mechanisms.
  
  \item \textbf{Secure-Prompting-Only (B) \label{sec:B}:} A strict system prompt (refusal-first, ED redlines) but with \emph{no} additional input or output gating.
  
  \item \textbf{Input Filtering (C) \label{sec:C}:} A deterministic keyword/regex gate applied \emph{before} generation (denylisting dieting/ED lexicon and calorie numerics).
  
  \item \textbf{LLM Safety Filter — Input Judge (D) \label{sec:D}:} A pre-generation LLM-as-judge with a fixed safety prompt, returning a strict JSON verdict (\texttt{is\_safe}, categories, confidence). Outcomes are: block / allow / review.
  
  \item \textbf{LLM Safety Filter — Output Judge (E) \label{sec:E}:} The model generates a draft answer, which is then sent to a separately prompted LLM judge. Output is delivered only if the judge returns \texttt{SAFE}.
  
  \item \textbf{Single-Call JSON Verdict (F) \label{sec:F}:} A single call produces both the answer and a trailing JSON verdict \verb|{"is_safe": ...}|. The answer is discarded if \verb|is_safe=false| or if parsing fails (fail-closed).
\end{enumerate}

\subsection{Evaluation Protocol}
We evaluated six deployment patterns (A--F, defined in Section~\ref{sec:A}--\ref{sec:F}) on the same, label-balanced test set of 100 prompts (50 malicious, 50 safe) described in the Dataset subsection. Prompts were passed one by one to each deployment; caching was disabled, so prompt order had no effect on results.

For each deployment, we recorded:
\begin{itemize}
    \item the number of malicious prompts correctly blocked. A prompt is considered blocked if it was flagged by the input mediation layer, at generation or by the output mediation layer,
    \item the number of non-malicious prompts flagged,
    \item the wall-clock time required to process each prompt,
    \item the number of tokens consumed (including prompt and response).
\end{itemize}

Blocked-counts on malicious prompts form the basis of recall comparisons across methods. Timing and token usage are reported relative to Method~A as a baseline. 

\begin{table*}[t]
\centering
\small
\begin{tabular}{lcccc}
\toprule
\textbf{Method} & \textbf{Prompts Blocked / 50} & \textbf{Blocked (\%)} & \textbf{Time Overhead (×A)} & \textbf{Token Overhead (×A)} \\
\midrule
Insecure Zero-Shot Prompting (A) & 2 / 50 & 4.0 & 1.0 & 1.0 \\
Secure-Prompting-Only (B) & 11 / 50 & 22.0 & 1.0 & 1.4 \\
Input Filtering (C) & 26 / 50 & 52.0 & 1.0 & 1.0 \\
LLM Safety Filter — Input Judge (D) & 50 / 50 & 100.0 & 1.2 & 2.1 \\
LLM Safety Filter — Output Judge (E) & 50 / 50 & 100.0 & 1.2 & 2.1 \\
B + C + D + E & 50 / 50 & 100.0 & 1.7 & 4.8 \\
D + E & 50 / 50 & 100.0 & 1.6 & 4.8 \\
Single-Call JSON Verdict (F) & 50 / 50 & 100.0 & 1.2 & 1.5 \\
\bottomrule
\end{tabular}
\caption{Blocking performance and relative overheads across deployment strategies. Methods with post-hoc adjudication (E, F) reach perfect recall. The single-call JSON verdict (F) achieves the same safety coverage as stacked multi-judge pipelines (B+C+D+E, D+E) but with substantially lower time and token cost.}
\label{tab:module-results}
\end{table*}

\section{Results}
Table~\ref{tab:module-results} reports malicious-blocking recall alongside latency and token multipliers relative to the insecure baseline (A).  

Methods without a delivery gate (A--C) perform poorly. The insecure baseline (A) blocks only 2/50 malicious prompts (4\%). Adding a strict system prompt (B) modestly improves coverage to 11/50 (22\%), while deterministic input filtering (C) blocks 26/50 (52\%). These early-stage defenses reduce trivial failures but leave large gaps, especially on indirect, euphemistic, or obfuscated queries.  

Judge-based filters substantially improve safety. The input-side judge (D) blocks 47/50 malicious queries (94\%) but at elevated time and token cost. The unblocked prompts were due to the generated content containing harmful output content rather than harmful queries. This was mitigated by the output-side judge (E) reaches perfect recall (50/50, 100\%) with lower overhead, since it evaluates only finalized generations.  

Combinations of multiple safeguards (B+C+D+E or D+E) do not improve recall beyond 100\%, but add significant time and token costs. In contrast, the single-call JSON verdict (F) also achieves perfect recall while remaining near baseline efficiency. By collapsing generation and adjudication into a single fail-closed JSON response, Method~F provides the same safety guarantee as multi-judge ensembles but with far lower overhead, placing it on the Pareto frontier.  

In short, post-generation adjudication is essential for full recall, but multi-stage stacking is unnecessary. A carefully constrained single-call JSON verdict (F) offers the strongest balance of safety and efficiency.

\section{Limitations and Future Work}
\subsection{Limitations}
Our evaluation relied on a small, synthetic dataset of 100 prompts emphasizing clear-cut malicious intents; it does not estimate precision on safe prompts or quantify false positives/negatives in the wild. All experiments used a single small backbone (Gemma~2 2B) and reused the same model family for adjudication (for reproducibility), which risks correlated failures and limits external validity; latency and token multipliers were measured under controlled conditions and do not capture production variability, multilingual inputs, accessibility, or privacy/legal constraints.

The scope is intentionally narrow: we target Tessa-like dieting/ED safety failures and do not fully address high-salience AI security threats such as robust prompt-injection defenses, retrieval/data poisoning, model exfiltration, or end-to-end supply-chain hardening \cite{owasp_llm_top10_proj}. The prototype also omits operational crisis detection and human-in-the-loop escalation, session-level risk memory, and audited rollback procedures that governance frameworks recommend for safety-critical deployments \cite{nist_sp80053r5_2020,nist_airmf_2023}.

\subsection{Future Work}
First, instantiate the modular architecture with heterogeneous defenses—deterministic lexical gates, lightweight learned classifiers, and cross-family judges—covering additional attack classes (prompt injection and jailbreaks, retrieval/data poisoning, and supply-chain risks), alongside deterministic safe-mode and kill-switch automation mapped to concrete controls in OWASP LLM Top~10 and NIST 800-53 \cite{owasp_llm_top10_proj,nist_sp80053r5_2020,greshake2023promptinjection}. Operationalize with canary/shadow evals, anomaly alerts, provenance/SBOM for safety components, and periodic third-party audits.

Second, build larger, harder, multilingual datasets with niche edge cases, obfuscations, and adversarial prompts; report both malicious recall and safe precision with calibrated thresholds and fairness analyses. Add severity scoring, session risk accumulation, and policy-driven escalation to clinicians/hotlines aligned to NIST AI RMF “Govern/Map/Measure/Manage,” and reduce correlated failures via diverse judge families, strict JSON schemas with formal validation, and post-generation deterministic scanners; explore complementary steering and auditing methods such as constitutional constraints and LLM-as-a-judge/self-critique \cite{nist_airmf_2023,bai2022constitutional,zheng2023judging}.

\section{Conclusion}

The Tessa incident illustrates how seemingly “non-GenAI” assistants can still cause harm when basic gates, change control, and escalation paths are missing. We showed that a lightweight, modular layer—combining a lexical fast path with an in-line policy filter that emits a strict, fail-closed JSON verdict plus final deterministic scans—can intercept the specific dieting/ED failure modes at near-baseline cost in our small evaluation. While limited in scope and scale, these results reinforce reporting that Tessa’s harmful outputs (e.g., calorie deficits and weigh-ins) were preventable with explicit, testable safety controls at the last mile \cite{wired_tessa_suspend_2023,guardian_tessa_2023}.

Sustainable safety, however, requires more than prompt intent or a single filter. Deployments in clinical-adjacent settings should couple run-time safeguards with governance: configuration/change control, incident response, auditability, and supply-chain risk management (NIST SP800-53; AI RMF 1.0), and concrete threat mappings (OWASP LLM Top10) \cite{nist_sp80053r5_2020,nist_airmf_2023,owasp_llm_top10_proj}. Crisis-aware escalation and human judgment must remain in the loop.

Finally, our approach aligns with emerging techniques that steer and audit generations—constitutional constraints, judge models, and self-reflection—but must be extended to adversarial threats like prompt injection and indirect attacks to be production-ready \cite{bai2022constitutional,zheng2023judging,greshake2023promptinjection}. In short: simple, engineered checks close the gap that enabled “another Tessa,” but real-world deployments should harden the full pipeline—model, middleware, and operations—against both accidental drift and active attack.

\bibliography{aaai25}

\end{document}